\documentclass[a4paper, 12pt]{nature1}
\usepackage{graphicx}
\usepackage{amssymb}
\usepackage[amssymb]{SIunits}

\newcommand{\ket}[1]{|{#1}\rangle}

\title{Strong interaction between light and a single
trapped atom without a cavity}

\author{Meng~Khoon Tey$^{\ref{add:nus}}$,
Zilong Chen$^{\ref{add:imre}}$, Syed~Abdullah Aljunid$^{\ref{add:nus}}$, Brenda
Chng$^{\ref{add:nus}}$, Florian Huber$^{\ref{add:munich}}$, Gleb
Maslennikov$^{\ref{add:nus}}$ \& Christian Kurtsiefer$^{\ref{add:nus}}$}

\begin{document}

\maketitle

\begin{affiliations}
\item \label{add:nus} Center for Quantum Technologies, National University of
  Singapore, 3 Science Drive 2, Singapore 117543.
\item \label{add:imre} Institute of Materials Research and Engineering, 3
Research Link,  Singapore 117602.
\item \label{add:munich} Department of Physics, Technical University of Munich,
James  Franck Street 85748 Garching, Germany.
\end{affiliations}


\begin{abstract}
Many quantum information processing protocols require efficient transfer of
quantum information from a flying photon to a stationary quantum system. To
transfer information, a photon must first be absorbed by the quantum
system. A flying photon can be absorbed by an atom residing in a high-finesse
cavity with a probability close to unity. However, it is unclear whether a
photon can be absorbed effectively by an atom in a free space. Here, we report
on an observation of substantial extinction of a light beam by a single
$^{87}$Rb atom through focusing light to a small spot with a single
lens. The measured extinction values are not influenced by
interference-related effects, and thus can be compared directly to the
predictions by existing free-space photon-atom coupling models. Our result
opens a new perspective on 
processing quantum information carried by light using atoms, and is important
for experiments that require strong absorption of single photons by an atom in
free space.  
\end{abstract}

Strong interaction between light and matter is essential for successful
operation of many quantum information protocols such as quantum
networking\cite{cirac:1997, duan:2001}, entanglement swapping between two
distant atoms \cite{monroe:2008, blinov:2004, volz:2006}, and implementation of
elementary quantum gates\cite{walls:1990}. These protocols consider quantum
states of localized carriers (nodes) like atoms, ions, or even atomic
ensembles, 
that exchange information through a quantum channel with help of ``flying''
qubits (photons). The quantum channels can be implemented via well-defined
photonic modes that couple the nodes with high efficiency. For example, in the
original proposal for quantum networks\cite{cirac:1997}, atoms
were placed in high-finesse cavities that not only provide a strong
interaction between a photon and an atom, but also 
ensure that most of the spontaneously emitted photons are collected into the
same mode. Experimental advances in atom-photon cavity QED indeed allowed
the information exchange between an atom and single photons in this
configuration to be carried out with high efficiency\cite{muller:1985,turchette:1995,rempe:2000,kimble:2007,gleyzes:2007}. However,
scaling such 
a scheme to many localized nodes is experimentally difficult, since
managing the losses and coupling of the intra-cavity field of high-Q cavities
to propagating modes of flying qubits is already quite challenging.

In an attempt to avoid the complications connected with cavities, one could
consider an interface between stationary and flying qubits in a simpler
free-space configuration, where the quantum channel is defined e.g. by a
Gaussian mode of a single mode optical fiber, and a single atom is strongly coupled to
this mode with help of a large numerical aperture lens.  Indeed, the common model describing the
interaction of a monochromatic plane wave with a two-level atom predicts a
scattering cross section of $\sigma=3\lambda^2/2\pi$. This area is close to a
diffraction limited spot size of a lens with a large numerical aperture, hence
suggesting a high coupling efficiency\cite{coup_eff} for such a system. On the other hand,  
for strong focusing where substantial coupling might be expected,
one has to carefully consider the electric field strength and
polarization within the focal `spot'\cite{vanenk:2001,vanenk:2004} because an
atom essentially interacts only with the field at its
location. The conclusion from such an attempt\cite{vanenk:2001} was that
for realistic lenses, only a low coupling efficiency
can be accomplished.
In view of those two contradicting opinions, we experimentally quantified the
coupling efficiency between a focused light beam and a single atom without a
cavity using a simple transmission measurement setup.

The first transmission spectrum of a single atom was observed for a $^{198}$Hg$^+$ ion\cite{wineland:1987}. There, the absorption
probability of the probe photons was estimated to be about
$2.5\times10^{-5}$. Recently perfomed experiments on single molecules and
semiconductor quantum dots\cite{gerhardt:2007,wrigge:arx,vamivakas:2007}
reported a signal contrast up to 13\%. However, these results do not reflect
the actual extinction of the excitation beam by the quantum systems
directly, since the signals observed were enhanced using the interference
between the light scattered by the single quantum systems and part of the
excitation light beam. 
 The main idea of our setup is to focus a weak and narrow bandwidth Gaussian 
light beam (probe) onto a single $^{87}$Rb atom using a lens. Part of the
probe is scattered by the atom. The remaining part is fully collected by a
second lens in the downstream direction, and delivered to a single photon
detector. Compared to the previous experiments, our setup allows us to
\textit{directly\/} measure the extinction of a probe beam by a single atom
(see methods) free of interference enhancement effects. The extinction value
obtained this way  sets a lower bound to the scattering probability of the
light by the atom (see methods).

\begin{figure}
\begin{center}
\includegraphics[width=0.9\columnwidth]{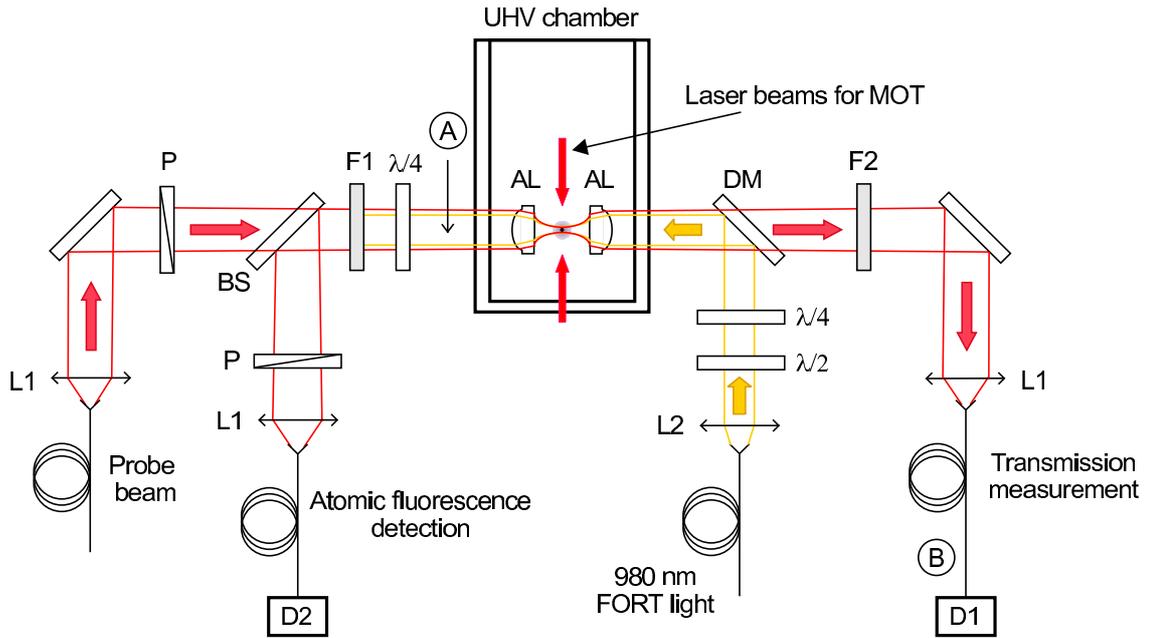}
\caption{\label{img:setup}
Experimental setup for measuring the extinction of a light beam by a single atom. AL: aspheric
lens ($f=4.5\,$mm, full ${\rm NA} =0.55$), P: polarizer, DM:
dichroic mirror, BS: beam splitter with 99\% reflectivity, $\lambda/4, \lambda/2$: quarter and half wave plates, F1: filters for blocking the 980 nm FORT light, F2: 
interference filter centered at $780\,$nm, D1 and D2: Si-avalanche
photodiodes. Four more laser beams forming the MOT lie in an orthogonal plane
and are not shown explicitly. 
}
\end{center}
\end{figure}

\begin{figure}
\begin{center}
\includegraphics[width=0.8\columnwidth]{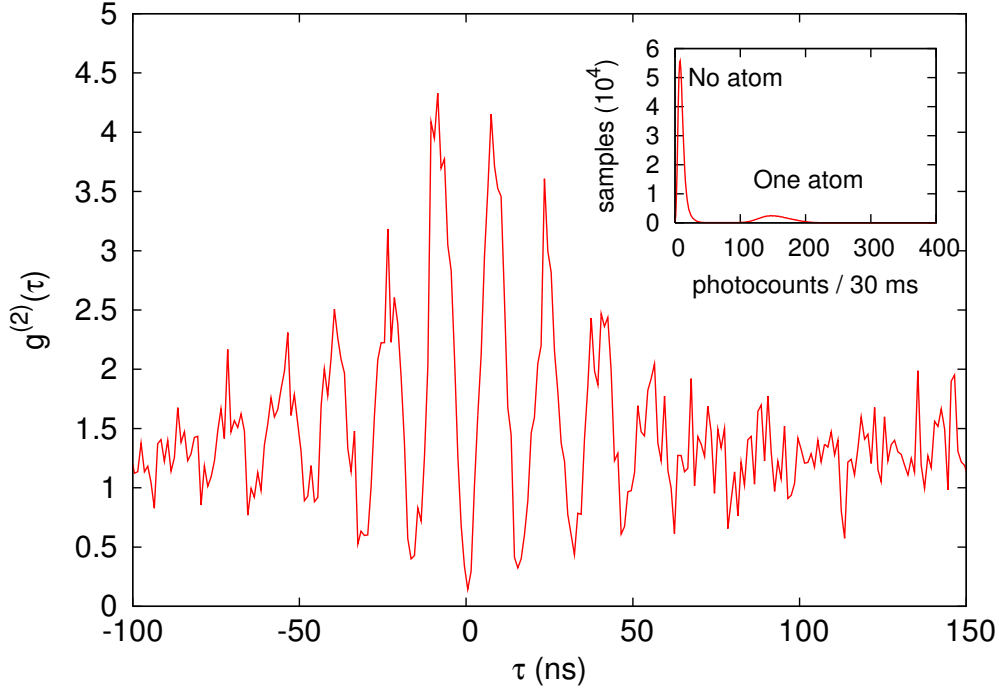}
\caption{\label{img:g2}
Normalized second-order correlation function versus time delay $\tau$ between
two photodetection events at detectors D1 and D2 (not corrected for background
counts) with clear anti-bunching at $\tau = 0$. The inset shows a histogram of
photocounts from the atomic fluorescence revealing the ``binary'' character of
the detected events due to collisional blockade\cite{grangier:2001}.
}
\end{center}
\end{figure}

Figure~\ref{img:setup} shows the schematic diagram of our experiment. The heart 
of the setup consists of two identical aspheric lenses (full ${\rm NA}=0.55,
f=4.5\,$mm), mounted in a confocal arrangement inside an ultra high
vacuum chamber. The Gaussian probe beam is first
delivered from a single mode fiber, focused by the first lens,
fully collected by the second lens, and finally coupled again into a single
mode fiber connected to a Si-avalanche photodiode. A $^{87}$Rb atom is trapped
at the focus between the two lenses by means of a far-off-resonant optical dipole
trap (FORT) formed by a light beam ($\lambda=980\,$nm) passing through
the same lenses. Cold atoms are loaded 
into the FORT from a magneto optical trap (MOT) surrounding the FORT. In this
experiment, the FORT beam has a waist of $1.4\,\mu$m at the focus\cite{comment:waist}. The maximal trapping potential at the center of the FORT
is about $h\cdot27\,$MHz. Due to the small size of the FORT, a collisional
blockade mechanism allows no more than one atom in the trap at any
time\cite{grangier:2001, schlosser:2002}. To confirm the single atom occupancy of the trap, we
extract the second order correlation function $g^{(2)}(\tau)$ from the
fluorescence of the trapped atom exposed to the MOT beams with the help of
detectors D1 and D2 that couple to the atom from opposite directions through
the same Gaussian mode (Fig. \ref{img:setup}). Figure \ref{img:g2} shows the histogram of the time delays
between photodetection events at detectors D1 and D2. It reveals a
Rabi oscillation with $\approx$\,62\,MHz and with a damping time compatible
with the spontaneous decay lifetime of the 5P state in $^{87}$Rb 
(27\,ns). An almost vanishing $g^{(2)}(\tau=0)$ indicates that no two
photons are emitted at the same time from the trap region, providing strong
evidence that we only have a single atom in the
trap\cite{diedrich:1987,weber:2006,meschede:1998}.
 The observation of a binary on/off fluorescence signal provides further evidence that there is either
one or no atom in the trap at any time\cite{grangier:2001}.

\begin{figure}
\begin{center}
\includegraphics[width=0.8\columnwidth]{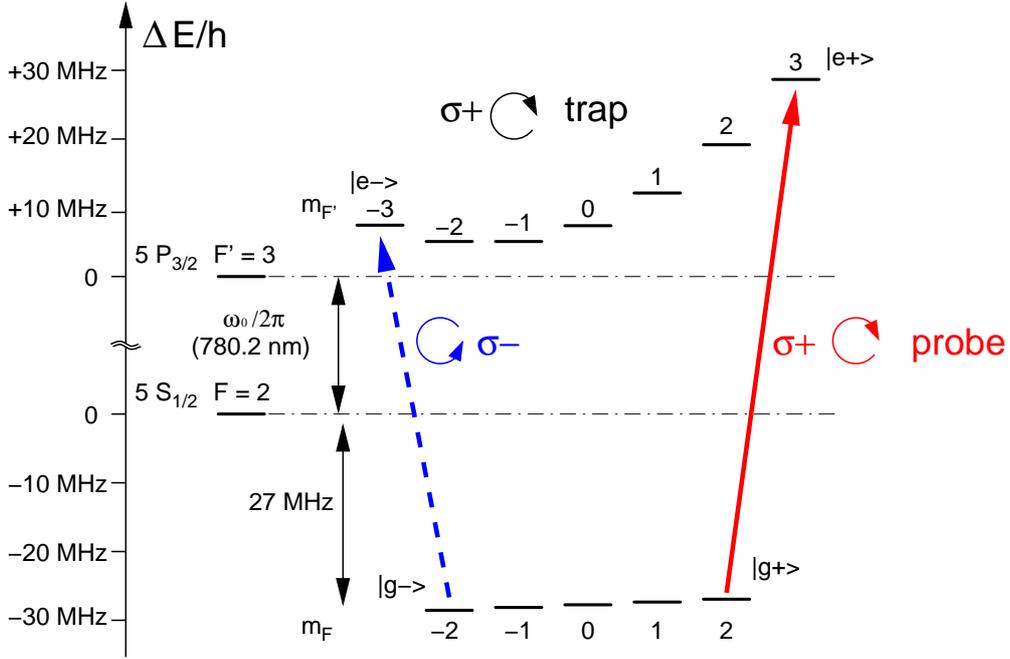}
\caption{\label{img:levelscheme}
Predicted AC Stark shift of a $^{87}$Rb atom in a circularly polarized FORT
for the parameters mentioned in the text. 
}
\end{center}
\end{figure}

We would expect to observe the largest extinction for a clean two-level system
with no other decay channels. Therefore, 
we use a circularly polarized probe beam that optically pumps the
$^{87}$Rb atom to a closed-cycling transition either
between $\ket{g+} = \ket{5{\rm S}_{1/2}, F=2, m_F =+2}$ and 
$\ket{e+} = \ket{5{\rm P}_{3/2},F'=3,m_{F'}=+3}$, or between
$\ket{g-} = \ket{F=2,m_F =-2}$ and $\ket{e-} = \ket{F'=3,m_{F'}=-3}$
(Fig. \ref{img:levelscheme}). As the MOT beams are turned off
during the measurement, the atom can be heated up 
and even kicked out of the FORT by the probe. To avoid this problem, the intensity of
the probe is reduced to a level where the actual photon scattering rate
was estimated to be around $2500\,$s$^{-1}$ (about
five times smaller than the longitudinal oscillation frequency of the atom in
the FORT). For such a low scattering rate, however, we need to ensure that the
atom does not leave the cycling transition between subsequent photon
scattering events. A magnetic field orthogonal to the quantization axis causes
the atom to undergo Larmor precession, leaking population from $\ket{g\pm}$ or
$\ket{e\pm}$ to other $\ket{m_F}$, $\ket{m_{F'}}$ states, which upsets the
clean two level system. To prevent this, we carefully zero the magnetic
field at the location of the trapped atom, and then apply a magnetic bias field
along the quantization axis during the measurement. Similarly, the
FORT-induced AC 
Stark shift breaks the degeneracy of the hyperfine states of the trapped
atom. If $\ket{g\pm}$ and $\ket{e\pm}$ (fixed through optical pumping by the
probe) 
are not the energy eigenstates of the atom in the FORT, population also leaks
out of the two-level system. Experimental evidence for
this was a reduction of the observed extinction by a factor of two for
linearly polarized FORT field. 
In our experiment, we therefore we adopt a circularly polarized FORT beam
counterpropagating with the circularly polarized probe.

Figure \ref{img:levelscheme} shows the calculated AC Stark
shift of the $5{\rm S}_{1/2}$, ${\rm F}=2$ and $5{\rm P}_{3/2}$, ${\rm F'}=3$
hyperfine states of the $^{87}$Rb atom under the influence of a circularly polarized
FORT light of 980 nm wavelength with a trapping potential depth of
$h\cdot27\,$MHz. The quantization axis of our system is chosen parallel to the
main propagation axes of the probe/FORT beams and such that the polarization of the FORT field is right hand circular. A
$\sigma^+$ probe refers to a circular polarization that drives the atom from
$\ket{g +}$ to $\ket{e +}$, and a $\sigma^-$ probe to one driving a $\ket{g
  -}$ to $\ket{e -}$ transition. At the center of the FORT, the energies of
$5{\rm  S}_{1/2}$ states are lowered by an average of $h\cdot27\,$MHz
(defining the trapping potential) with a small sublevel energy splitting of
$\approx1$\,MHz. The $5{\rm P}_{3/2}$ levels shift upwards and
are strongly split, forming a repulsive potential. The resulting shifts of the
resonance frequency for different transitions can be observed directly in a
transmission measurement in which the frequency of the probe is scanned over
the resonance frequency of the trapped atom.

\begin{figure}
\begin{center}
\includegraphics[width=0.8\columnwidth]{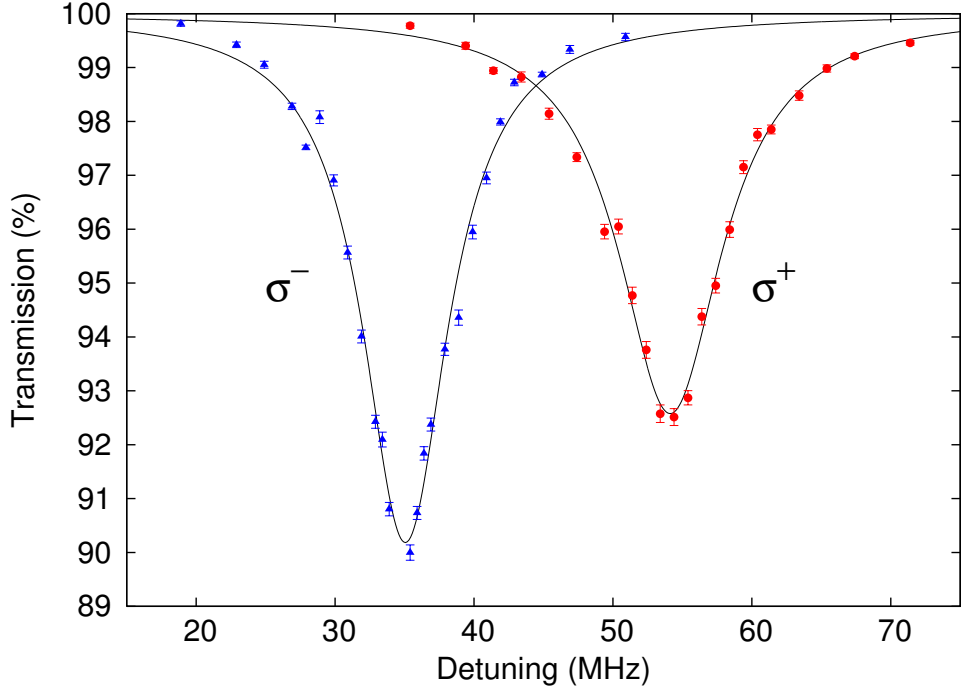}
\caption{\label{img:transmission}
Transmission of the probe beam versus detuning from the natural
resonant frequency $\omega_0/2\pi$ of the $\ket{g}$ to $\ket{e}$ transition. The
absolute photon scattering rate is kept at $\approx2500\,{\rm s}^{-1}$ for
every point by adjusting the probe intensity according to the measured
extinction. The solid lines are Lorentzian fits. 
}
\end{center}\end{figure}

Figure \ref{img:transmission} shows the transmission of the probe as a
function of detuning from the natural resonant frequency
$\omega_0/2\pi$ of the $\ket{g}$ to $\ket{e}$ transition (see methods for
transmission measurement procedures). The two spectra of a single
$^{87}$Rb atom were obtained for $\sigma^+$ and $\sigma^-$ probes, while 
keeping the handedness of the FORT beam fixed. As expected, the atomic
resonance frequency is different for the two probe polarizations, and agrees
very well with prediction shown in Figure \ref{img:levelscheme}. 
The Lorentzian fit to the transmission spectrum for the $\sigma^-$ probe shows a 
maximum extinction of $9.8 \pm 0.2 \%$ with a full-width-half-maximum (FWHM)
of $7.5 \pm 0.2\,$MHz. The $\sigma^+$ probe gives
a maximum extinction of $7.4 \pm 0.1 \%$ with a FWHM of $9.1 \pm
0.3\,$MHz. From the fact that the D2 transition of $^{87}$Rb has a
natural linewidth of 6.0\,MHz and that the linewidth/stability of
the probe laser is about 1\,MHz, we conclude that an atom exposed
to the $\sigma^-$ probe has been successfully kept in a two-level cycling
transition, and it experiences very small spectral broadening caused by
position dependent AC Stark shift in the FORT. However, the same conclusion
cannot be made for the atom exposed to a  $\sigma^+$ probe. A possible
explanation is that optical pumping by the $\sigma^+$
probe is less effective because a probe frequency resonant to the
$\ket{g+}$ to $\ket{e+}$ transition is further detuned from the resonant
frequencies of other $\ket{F~=2,~m_F}$ to $\ket{F'~=~3,~m_{F'}}$ transitions,
whereas the resonance frequency of $\ket{g-}$ to $\ket{e-}$ is less detuned from
other transitions (Fig. \ref{img:levelscheme}). Furthermore, a FORT wavelength of 980\,nm
forms a
repulsive potential for the 5P$_{3/2}$ levels of the $^{87}$Rb atom. As the
energy of $\ket{e+}$ is higher than that of the $\ket{e-}$, an atom in 
$\ket{e+}$ experiences a stronger repulsive force from the FORT on average. As
a result, a trapped $^{87}$Rb atom might be more susceptible to increase of
kinetic energy under the $\sigma^+$ probe, and thus oscillates more strongly around
the focus. 

Coming back to the photon-atom coupling efficiency, we want to emphasize that
an extinction of 9.8\% observed for a probe focused to $\approx860\,$nm
waist\cite{comment:waist} is large when compared to results reported from
experiments performed on single
molecules and quantum dots\cite{gerhardt:2007,wrigge:arx,vamivakas:2007}. There,
the excitation light field was either confined with a small
aperture of $\approx100\,$nm\cite{gerhardt:2007}, or focused by using solid immersion
lenses\cite{wrigge:arx,vamivakas:2007} that provide much tighter focusing than
in our case. In all these experiments quantum systems were embedded into
complex solid state host environments which complicates the theoretical
treatment of light scattering. 
The conceptual simplicity of the system we investigate and the fact that we
directly measure the extinction of the probe beam allows a clean comparison
with existing photon-atom coupling
models\cite{vanenk:2001,vanenk:2004,sondermann:2007}.

One of the models that closely describes our experiment was
presented by van Enk and Kimble \cite{vanenk:2001}. It considers a
monochromatic and circularly polarized Gaussian beam focused by an ideal thin
lens onto a two-level atomic system. Estimations based on that model gave a
very dim outlook on the effectiveness of coupling light to an atom using a
lens. In particular, a direct application of the method described there
predicts a maximum scattering probability of 2.2\% for our experimental
parameters. As it turns out, two approximations adopted in the model
(parabolic wave front after the lens, and no change to the
polarization of a light beam passing through the lens) has greatly
underestimated scattering 
probability for stronger focusing. Dropping these approximations, we find (with
otherwise same methods) a scattering probability of 20.3\% for our
experimental parameters\cite{mk:2008}. The residual difference between the
predicted and
measured values could be both due to the imperfections of our aspheric lens,
and the fact that the atom is not completely stationary at the
focus. Applying this model for an even tighter focus, a very high scattering
probability of up to 95\% is predicted (for focusing NA$\approx0.9$)
\cite{mk:2008}. Such a high 
scattering probability is at odd with other photon-atom coupling
models which suggest a maximum scattering probability of 50\% for a light beam
focused by a lens as in our setup\cite{vanenk:2004,sondermann:2007}; further
experimental work is required to check this discrepancy.

In conclusion, we experimentally observed a substantial extinction of a weak
coherent light field by a single atom by focusing the light beam using a
lens. In
particular, a coupling efficiency of at least 9.8\% has been achieved with a
focused beam waist of $\approx0.86\,\mu$m. Such values might appear to be
small compared to the maximum achievable with the help of a cavity. In
practice, however, due to mode-matching issues and other passive losses,
achieving very good coupling of light into a high finesse cavity is
nontrivial. This problem reduces the overall photon-atom coupling
efficiency between a truly 'flying' qubit and an atom when using a
cavity\cite{kimble:2007}. Contrary, a lens system suffers much less
from reflection losses. This advantage, together with the simplicity of such
configuration would make such a photon-atom coupling scheme very appealing to
many applications involving quantum state transfer from photons to
atoms. Furthermore, the strong interaction of the atom with a flying qubit
suggests using the atom as a mediator for a photon-photon interactions,
pointing in a new direction for implementing photonic quantum gates.

\begin{methods}
\section*{Direct extinction measurements}
In general, extinction is obtained by comparing the transmitted power of the
probe with and without the sample in the optical path of the probe. In usual
extinction measurements, e.g. as implemented in a commercial
spectrophotometer, the probe beam is collected fully by the power measuring
device. However, this is not the case in the extinction measurements on single
quantum systems reported so far, e.g. in\cite{gerhardt:2007,wrigge:arx,vamivakas:2007}.
The reason is that substantial extinction of a probe beam by single quantum systems generally
requires strong focusing. It is, nevertheless, difficult if
not impossible in most experiments to collect the strongly diverging probe
fully after the focus. As such, the `extinction' measured in such experiments
is not the extinction in the usual sense and cannot be used in a
straightforward way to quantify the actual scattering probability of the probe
by the quantum system without further model assumptions. In our experiment, we
collect all of the diverging probe light, and thus are able to carry out a
\textit{direct} extinction measurement.

The measured transmission $T$ is related to the scattering probability
$P_{sc}$ by $T=1-P_{sc}+\alpha P_{sc}$, where $\alpha$ represents the
percentage of scattered light collected by the transmitted power detector. The
extinction $\epsilon=1-T$ is thus related to the scattering
probability by $P_{sc}=\epsilon/(1-\alpha)$.  The collection efficiency
$\alpha$ in this experiment is estimated to be less 5\%, so $\displaystyle
P_{sc}\approx \epsilon$.

\section*{Losses and interference artefacts}
We carefully quantified the losses in the transmission channel to make
sure our results do not suffer from interference artefacts (interference
between partially collected probe and scattered light can lead to value of
'extinction' larger than the scattering probability). The total transmission from
point A in Fig. \ref{img:setup} (before the vacuum chamber) to point B (after
the single-mode fiber and just before the detector) is 53\%. The 47\% loss
include 21.6\% loss from the four uncoated window surfaces of the vacuum chamber and
the two aspheric lenses; 5.3\% loss over two dichroic mirrors, an interference
filter (peak transmission $T$=96\% at 780\,nm) and a mirror; and 28.4\% 
coupling loss into an uncoated single mode fiber. All the losses are caused by reflection except for 20\% loss at the fiber coupling that is due to imperfect mode matching. Since the scattered field and the probe field should experience the same reflection loss at each surface, we are reasonably confident that our results are free
from interference artefacts.

\section*{Sequence for transmission measurement}
Once an atom is loaded into the FORT, it triggers the transmission measurement sequence. The main steps include: step 
1. Switching off the MOT beams and the MOT quadrupole coil current; step 2. Application of a magnetic bias field of
$\approx\!2$\,G along the quantization  axis; step 3. Waiting for $20\,$ms so that current in the coils
stabilizes and optically pumping the atom into either $\ket{g+}$ or $\ket{g-}$ at the same time; step 4. Recording the photo counts $n_m$ of the transmitted
 probe beam for $\tau_m$ ranging from 130 to 140\,ms with detector D1;
 step 5. Switching on the MOT beams to check whether the atom
  is still in the FORT by monitoring fluorescence with detector D2; if
``yes'', turn off the MOT beams and repeat step 3 and 4; 
step 6. Otherwise, recording the photo counts $n_r$
of the transmitted probe beam with detector D1 for $\tau_r=2\,$s without an
  atom in the trap for reference; step 7. Turning on the MOT beams and the quadrupole coil current, waiting for
  another atom to be loaded in the FORT. 

A transmission value $T$ is obtained for each atom trapping event by
$\displaystyle T=\frac{\sum n_m}{\sum \tau_m}\frac{\tau_r}{n_r}$, where the
summation is carried over all contiguous measurement intervals $m$ for which
an atom was found in the trap. The average time an atom stays in the
FORT is about 1.5\,s. A single data point in figure \ref{img:transmission} is the
average of about 100 of such transmission values, each weighted by
$\displaystyle \frac{\tau_r\sum{\tau_m}}{\tau_r+\sum{\tau_m}}$.

The error is dominated by photo counting shot noise, our error bars indicate
$\pm1$ standard deviation. During the transmission measurement process, the
atom may fall into the $\ket{5S_{1/2},F=1}$
metastable ground state, which is off resonant with the probe. To bring it
back to the pumping cycle, circularly polarized light resonant with the D1
transition is mixed into the probe beam, and later
removed with an interference filter F2 (Fig. \ref{img:setup}). 

\end{methods}

\begin{addendum}
\item We would like to acknowledge helpful discussions with Valerio
  Scarani. This work was partially supported by the Singapore Ministry of Education
  under FRC grant R-144-000-174-112.
\item[Correspondence] Correspondence and requests for materials should be addressed to C.K. (email: christian.kurtsiefer@gmail.com).
\end{addendum}


\begin{thebibliography}{10}
\bibitem{coup_eff}
The coupling efficiency refers, here, to the absorption probability of a flying photon by a stationary quantum system. For a free atom, the absorption probability is also equivalent to the scattering probability of the photon by the atom.
\bibitem{comment:waist}
Focal waists are inferred from the measured input waist using paraxial
approximation.
\end{thebibliography}
\end{document}